\documentclass{article}
\usepackage{epsfig}

\title{\bf\Large {ITINERANT FERROMAGNETISM AND SUPERCONDUCTIVITY}}
\author{Naoum Karchev\cite{byline}}
\begin{document}

%
\maketitle  

Superconductivity has again become a challenge following the discovery of
unconventional superconductivity. Resistance-free currents have been 
observed in heavy-fermion materials, organic conductors and copper oxides.
The discovery of superconductivity in a single crystal of $UGe_2$, $ZrZn_2$
and $URhGe$ revived the interest in the coexistence of superconductivity
and ferromagnetism. The experiments indicate that: i)The superconductivity
is confined to the ferromagnetic phase. ii)The ferromagnetic order is
stable within the superconducting phase (neutron scattering experiments). iii)The 
specific heat anomaly associated with the superconductivity
in these materials appears to be absent. The specific heat depends on the
temperature linearly at low temperature.

I present a review of the recent experimental results and the basic theoretical
ideas concerning ferromagnetic superconductivity (FM-superconductivity) induced
by ferromagnetic spin fluctuations. A particular attention is paid to the magnon
exchange mechanism of FM-superconductivity.

 \section{\bf Introduction}

The discovery of unconventional superconductivity caused an explosive growth 
of activities in various fields of condensed-matter research, stimulating not
only studies of the basic mechanisms leading to this phenomenon, but also a widespread
search for new technological applications. Resistance-free currents have been 
observed in heavy-fermion materials $CeCu_2Si_2$\cite{ns1}, $UBe_{13}$\cite{ns2},
$UPt_3$\cite{ns3}, and $U_{1-x}Th_xBe_{13}$\cite{ns4}, in organic conductors
\cite{ns5,ns6,ns7}, copper oxides\cite{ns8,ns9} and layered ruthenate 
$Sr_2RuO_4$\cite{ns10}. The great interest in 
the unconventional superconductivity is in particular due to their rather
different normal- and superconducting-state properties. The mechanism of superconductivity
and symmetry of the order parameter are the main puzzling of on-going research.

The Cooper pairing of conducting electrons is characterized by the gap function
$\Delta_{\sigma,\sigma'}(\textbf{p})$, which is a $2\times 2$ matrix-valued function of 
the wave vector $\textbf{p}$. There is a symmetry relation which follows from the 
anti-commutation of spin $\frac 12$ fermions
\begin{equation}
\Delta_{\sigma,\sigma'}(\textbf{p})\,=\,-\,\Delta_{\sigma',\sigma}(-\textbf{p}),
\label{ns1}
\end{equation}

The inversion symmetry is expressed by
\begin{equation}
P\Delta_{\sigma,\sigma'}(\textbf{p})\,=\,\Delta_{\sigma,\sigma'}(-\textbf{p}),
\label{ns2}
\end{equation}
and the time reversal symmetry by
\begin{eqnarray}
T\Delta_{\sigma,\sigma}(\textbf{p}) & = & \Delta^\ast_{-\sigma,-\sigma}(-\textbf{p}),\,
{\textit{where}}\,\, (\sigma)=(\uparrow,\downarrow)~
\textit{and}\,\, (-\sigma)=(\downarrow,\uparrow) \nonumber \\
T\Delta_{\sigma,\sigma'}(\textbf{p}) & = & -\Delta^\ast_{\sigma',\sigma}(-\textbf{p}), \textit{when}\,\,
 \sigma\neq\sigma'. 
\label{ns3}
\end{eqnarray}

One can represent the gap matrix in the form
\begin{equation}
\Delta_{\sigma,\sigma'}(\textbf{p})\,=\,i\Delta(\textbf{p})(\tau_2)_{\sigma,\sigma'}\,+\,
id_{\mu}(\textbf{p})(\tau_{\mu}\tau_2)_{\sigma,\sigma'}.
\label{ns4}
\end{equation}
where $\tau_1,\tau_2$ and $\tau_3$ are the Pauli matrices, $\Delta(\textbf{p})$ is spin singlet function 
and $d_{\mu}(\textbf{p})$ is spin triplet one. The singlet part of the gap is a symmetric function
$\Delta(-\textbf{p})\,=\,\Delta(\textbf{p})$, while the triplet part is an antisymmetric one 
$d_{\mu}(-\textbf{p})\,=\,-d_{\mu}(\textbf{p})$.

In conventional superconductors, the quasi-particles  form Cooper pairs in a spin-singlet state 
$\Delta(\textbf{p})\neq0$, $d_{\mu}(\textbf{p})=0$, which has zero total spin. The existence of the gap in
the quasi-particle spectrum leads to unusual thermodynamic properties of the systems: 
i) The specific heat decreases exponentially as $\exp{(-\Delta(0)/k_BT)}$ at low temperature, 
as opposed to the linear temperature dependence in the Fermi liquid theory\cite{ns11}. 
ii) The same anomalous temperature 
dependence shows the paramagnetic susceptibility in s-type superconductors\cite{ns12}.
In the case of the singlet s-pairing, the nuclear spin-lattice relaxation rate $1/T_1$ exhibits 
a peak just below the superconducting transition temperature\cite{ns13}. 
Finally, the time-reversal and parity symmetries are not broken in conventional superconductors.
All these properties are well understood on the basis of the Bardeen, Cooper, Schrieffer (BCS) theory
of superconductivity\cite{ns14}

Alternatively, the quasi-particles can form Cooper pairs in a spin-triplet state $\Delta(\textbf{p})=0$, $d_{\mu}(\textbf{p})\neq0$, analogous to the "p-wave" state of paired neutral fermions in superfluid 
${}^3He$\cite{ns15}. At present, the heavy fermion compound $UPt_3$ and layered ruthenate 
$Sr_2RuO_4$ are the only known spin-triplet superconductors. The most direct way to identify
the spin state of Cooper pairs is from measurements of their spin susceptibility, which can be determined
by the Knight shift\cite{ns16}, and from measurements of nuclear
spin-lattice relaxation rate $1/T_1$, 
probed by nuclear magnetic resonance and nuclear quadrupole resonance. No change in spin susceptibility 
was observed on passing through the 
superconducting transition in layered perovskite $Sr_2RuO_4$\cite{ns16}, and 
$UPt_3$\cite{ns17,ns18}. The relaxation rate $1/T_1$ measured in $Sr_2RuO_4$\cite{ns19} did not show coherence 
peak. Muon spin-relaxation measurements, for the same materials, reveal the spontaneous 
appearance of an internal magnetic field below the transition temperature. The appearance of such a field
indicates that superconducting state is characterized by the breaking of time-reversal symmetry\cite{ns20}. 
The unconventional nature of the superconductivity in $UPt_3$ is confirmed by the observation of three 
superconducting phases denoted as A,B, and C in the $H$(magnetic field)-$T$(temperature) phase diagram. Phases
meet each other at a tetracritical point\cite{ns21}.   
 
On theoretical ground, the unconventional superconductivity was described as a spin-triplet superconductivity: non-unitary spin-triplet superconductivity in $UPt_3$\cite{ns22}, and odd pairing state, which is 
two-dimensional analogue of the Balian-Werthamer state of ${}^3He$, in $Sr_2RuO_4$\cite{ns23}. The non-unitary spin-
triplet is defined by spin-1 complex vectorial function $\textbf{d}(\textbf{p})$ which satisfy 
$\textbf{d}^\ast (\textbf{p})\times\textbf{d}(\textbf{p})\neq 0$. 
A test for odd-triplet pairing was proposed by T.M.Rice and M.Sigrist\cite{ns23}. They consider a sandwich of thin film of $Sr_2RuO_4$ between two singlet superconductors with higher transition temperature. Above $T_{sc}$ of $Sr_2RuO_4$ this system should behave like a standard SNS Josephson junction where the coupling is due to proximity-induced singlet pairs in $Sr_2RuO_4$. Below $T_{sc}$, however, the Josephson coupling should decrease because as triplet pairing appears in $Sr_2RuO_4$ the proximity-induced singlet pairing will be suppressed. The anomalous temperature dependence of the Josephson effect would confirm the odd-parity symmetry of the order 
parameter in $Sr_2RuO_4$.

The discovery of superconductivity in a single crystal of $UGe_2$\cite{ns24}, $ZrZn_2$\cite{ns25}
and $URhGe$\cite{ns26} revived the interest in the coexistence of superconductivity
and ferromagnetism. The experiments indicate that: i)The superconductivity
is confined to the ferromagnetic phase. ii)The ferromagnetic order is
stable within the superconducting phase (neutron scattering experiments).
iii)The specific heat anomaly associated with the superconductivity
in these materials appears to be absent. The specific heat depends on the
temperature linearly at low temperature.

The interplay of superconductivity and magnetism has a long history. The possible
coexistence of superconductivity and ferromagnetism was considered as a theoretical
possibility for weak itinerant ferromagnets many years ago\cite{ns27}. In systems
like $ErRh_4B_4$\cite{ns28a,ns28b} and $HoMo_6S_8$\cite{ns28c}, 
it was observed that s-wave superconductivity
gives way to ferromagnetism at intermediate temperatures. In all these
compounds, superconductivity and magnetism originate from different part of the
electron system. 
In contrast, in $UGe_2$, $ZrZn_2$, and $URhGe$ apparently the same 
band electrons are subject to the ferromagnetic and superconducting instability.    

I present a review of the recent experimental results and the basic theoretical
ideas concerning ferromagnetic superconductivity (FM-superconductivity) induced
by ferromagnetic spin fluctuations. The review should be considered as a "progress 
report" in which I have attempted to focus on some basic aspects of this rapidly
evolving field. Another goal of the review is to provide the reader with a simple
overview of the experimental situation in ferromagnetic superconductors, summarizing 
the basic agreements and disagreements between theory and experiment.                
For additional literature the reader should consult other review articles\cite{ns29}.

The review is organized as follows. In the next section, a review of the experiments 
concerning  $UGe_2$, $ZrZn_2$ and $URhGe$ compounds is given. 
Theoretical models to describe superconductivity induced by spin fluctuations are 
presented in the third section. A particular attention 
is paid to the magnon exchange mechanism of FM-superconductivity.

\section{\bf Ferromagnetic superconductivity-experimental status}    

The coexistence of superconductivity and ferromagnetism has been studied for many years,
but it has only recently been demonstrated to occur experimentally\cite{ns24,ns25,ns26}.
The most surprising fact is that the superconductivity occurs only in the ferromagnetic
phase. 
The second surprise is that the specific heat anomaly associated with the superconductivity 
in these materials appears to be absent. 

Recent experiments on single crystal\cite{ns30} indicate that $UGe_2$ has the base-centered
orthorhombic crystal structure $(Cmmm)$ with zigzag chains of nearest-neighbor uranium ions. 
The structure is shown in Fig.1 taken from Ref.\cite{ns31}. 
\begin{center}
\begin{figure}[htb]
\label{nsfig1}
\centerline{\psfig{file=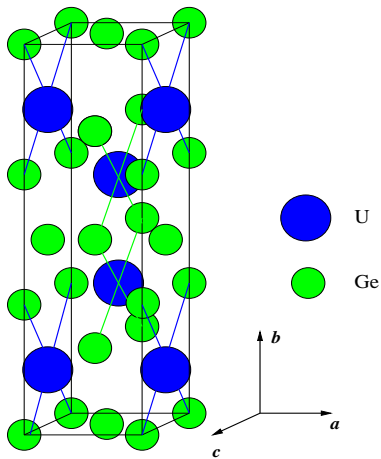,width=5cm,height=6cm}}
\caption{The base-centered orthorhombic $Cmmm$ crystal structure of $UGe_2$.}
\end{figure}
\end{center}

\vspace*{-0.7cm}

In the uranium
compounds known as "heavy-fermion systems" the $5f$ electrons are highly localized and interact 
with fermions from more delocalized levels, giving rise to fermion excitations characterized
by large effective masses. In $UGe_2$ , however, the $5f$ electrons are more itinerant than in
many heavy-fermion systems. Specific heat measurements show that $\gamma$ coefficient 
$\gamma=C(T)/T$ is about 10 times smaller than in conventional heavy-fermion 
$U$-compounds($\gamma\approx 35 mJ/K^2$\cite{ns32}), which suggests that these electrons behave 
more like the $3d$ electrons in the traditional itinerant ferromagnets such as $Fe$, $Co$ and $Ni$.
$UGe_2$ differs from the $3d$ metals mainly in having a stronger spin orbit interaction that leads
to an unusually large magnetocrystalline anisotropy with easy magnetization axis along $\hat{a}$
(shortest crystallographic axis, Fig.1). 

At ambient pressure $UGe_2$ is an itinerant ferromagnet below the Curie
temperature
 $T_c=52K$, with low-temperature ordered moment of $\mu_s=1.4\mu_B/U$. 
With increasing pressure the system passes through two
successive quantum phase transition, from ferromagnetism to
FM-superconductivity at $P\sim 10$ kbar, and at higher pressure 
$P_c\sim$ 16 kbar to paramagnetism\cite{ns24,ns33}. The resulting 
pressure $p$ temperature $T$ phase diagram for $UGe_2$ is shown in Fig.2(a)
taken from Ref.\cite{ns33}.
\begin{center}
\begin{figure}[htb]
\label{nsfig2}
\centerline{\psfig{file=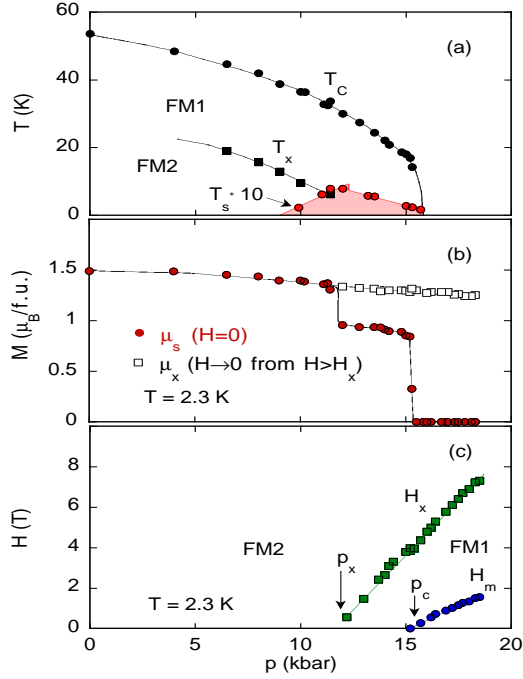,width=7cm,height=9cm}}
\caption{(a) The $p-T$ phase diagram of $UGe_2$. $T_c$ is the Curie temperature
$T_s$ is the superconducting temperature.(b) Pressure dependence of $\mu$ at
$T=2.3K$(full circles). The moment obtained by extrapolating the data from 
above $H_x$ to zero (squares). (c) Pressure dependence of the fields $H_x$ and
$H_m$ of metamagnetic transitions (at which $dM/dH$ has a local maximum) at 
$T=2.3K$}    
\end{figure}
\end{center}

\vspace*{-0.7cm}

At the pressure where the superconducting transition temperature 
is a maximum $T_{sc}=0.8K$, 
the ferromagnetic state is still stable with $T_c=32K$, and 
the system undergoes a first order metamagnetic transition between two
ferromagnetic phases $FM2\rightarrow FM1$ with different ordered moments 
\cite{ns34}. 
The pressure dependence of the moment $\mu$ at $2.3K$ is shown in Fig.2(b)
taken from Ref.\cite{ns34}. (The symbol $M$ is used for the magnetization 
in an external field).      

The survival of bulk ferromagnetism below $T_{sc}$ has been confirmed 
directly via elastic neutron scattering measurements\cite{ns33}.
The specific heat coefficient $\gamma=C/T$ increases
steeply near 11 kbar and retains a large and nearly constant value\cite{ns35}. 

The resistivity measurements reveal\cite{ns33} the presence of an additional
phase line that lies entirely within the ferromagnetic phase. It is suggested by
a strong anomaly seen in the resistivity. The characteristic temperature of
this transition, $T_x(p)$, decreases with pressure and disappears at a pressure
$p_x$ close to the pressure at which the superconductivity is strongest (Fig2(a)). 
The additional phase transition demonstrates itself and through the change in 
the $T$ dependence of $\mu(T)$\cite{ns34,ns35}.
In Fig.3, taken from\cite{ns34}, the temperature dependence of the ordered magnetic
moment, is shown. A clear change in the $T$ dependence of $\mu(T)$ occurs at $T_x$.  
\begin{center}
\begin{figure}[htb]
\label{nsfig3}
\centerline{\psfig{file=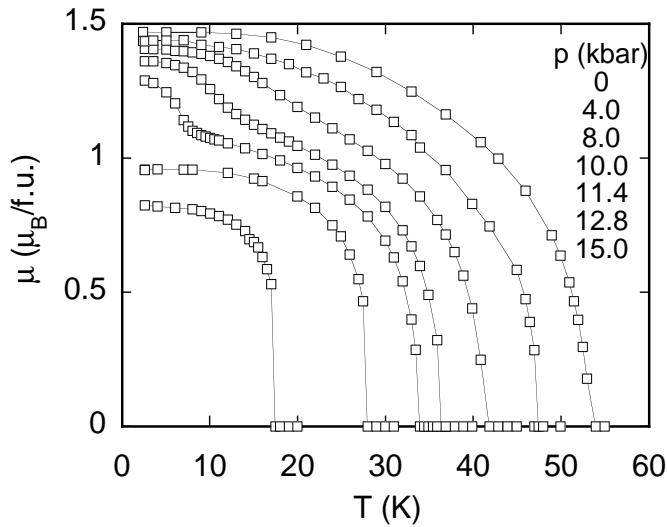,width=9cm,height=7cm}}
\caption{Temperature dependence of the ordered ferromagnetic moment.
Curves correspond from top to bottom to the pressures indicated in the
top right corner of the frame.}
\end{figure}
\end{center} 

\vspace*{-0.7cm} 

The field dependence of magnetization at $2.3K$ for different $p$ is shown in Fig.4,
taken from\cite{ns34}.
For pressure $p>p_x$ a large increase in the magnetization is observed at a field $H_x$.
It is defined as the field at which $dM/dH$ has a local maximum. It is plotted as a 
function of pressure in Fig.2(c). For $p>p_c$ the magnetization undergoes a second 
increase at low field $H_m$ corresponding to the paramagnetic$\rightarrow$ ferromagnetic 
transition.

It is important to stress that the longitudinal uniform susceptibility, given by the slope
$dM/dH$ at $H=0$, is small at a pressure $p_x$ close to the pressure at which the 
superconductivity is strongest, and increases rapidly with increasing of $p$, while the
superconducting transition temperature decreases. This experimental observation
suggests that the longitudinal spin fluctuations suppress the formation of Cooper pairs.  

\begin{center}
\begin{figure}[htb]
\label{nsfig4}
\centerline{\psfig{file=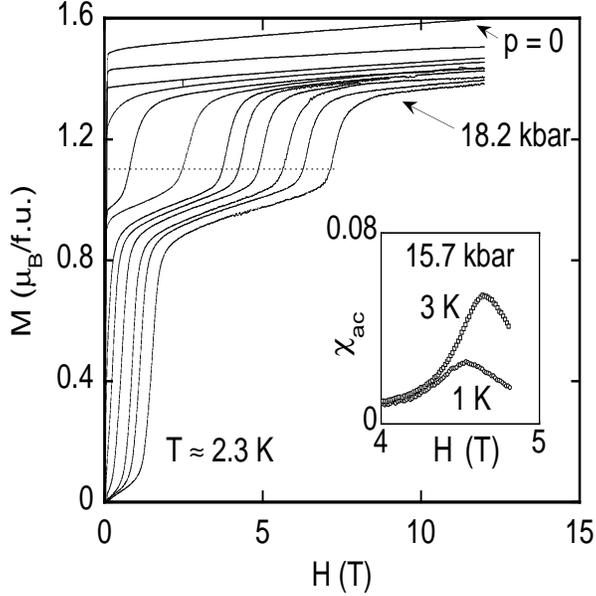,width=8cm,height=8cm}}
\caption{The field dependence of the easy-axis magnetization at 2.3K. Curves correspond from
top to bottom to p=0,6.5,9.0,11.1,12.8,15.3,15.5,16.0,16.7,17.3, and 18.2 kbar. The inset 
shows the ac susceptibility.}
\end{figure}
\end{center}

\vspace*{-0.7cm}
 
The anomaly at $T_x$ is quite similar to that observed in $\alpha$ uranium. 
For $\alpha$ uranium, there is direct evidence that the anomalies are due to the formation of
charge density wave (CDW), resulting from nesting at the Fermi surface. 
Although band structure calculations\cite{ns31,ns36} indicate that a spin-majority Fermi
surface sheet could become nested as a function of the magnetic polarization, no direct 
evidence for a charge density or spin density wave has been found in neutron 
diffraction studies\cite{ns37}. 

The studies of poly-crystalline samples of $UGe_2$ show that T-P phase diagram is very similar
to those of single-crystal specimens of $UGe_2$\cite{ns38}. This result suggests that high-purity
specimens with long mean free paths are not necessary, at least in the case of $UGe_2$, in order
to observe superconductivity. 

It is much less known about the $ZrZn_2$ and $URhGe$ compounds. 
The ferromagnets $ZrZn_2$ and $URhGe$ are superconducting at ambient pressure 
with superconducting critical temperatures $T_{sc}=0.29K$\cite{ns25} and 
$T_{sc}=0.25K$\cite{ns26} respectively.
$ZrZn_2$ is ferromagnetic below the Curie temperature $T_c=28.5K$ with 
low-temperature ordered moment of $\mu_s=0.17\mu_B$ per formula unit, while 
for $URhGe$\,\,$T_c=9.5K$ and $\mu_s=0.42\mu_B$. The quadratic low-temperature
dependence of the squared magnetization is characteristic of simple itinerant
ferromagnetism\cite{ns39}.  
The low Curie temperatures and small ordered moments
indicate that compounds are close to a ferromagnetic quantum critical point.

The physical properties of $URhGe$ at zero pressure closely resemble those
of $UGe_2$ at the high pressures where superconductivity is found. Although 
the space groups and detailed structures of $URhGe$ and $UGe_2$ are different,
both structures are orthorhombic and contain zigzag chains of nearest-neighbor 
uranium ions. Neutron scattering experiments reveal that the magnetization is
almost entirely attributable to uranium $5f$ electrons.   

The observation of a large jump in the specific heat, at the same temperature 
as the onset of superconductivity, demonstrates that the transition to 
superconductivity in $URhGe$ is a bulk phase transition. It also shows that
superconductivity involves the same itinerant electrons that are responsible for
ferromagnetism. At low temperature the specific
heat coefficient $\gamma$ is twice as smaller as in the ferromagnetic phase.

The superconductivity in $ZrZn_2$ has a number of remarkable features. 
First, it only appears to occur in high-purity single-crystal sample. Second, there
is no superconducting anomaly in the specific heat. It means that the superconducting 
state is strongly gapless with large portion of the Fermi surface, or even all of it,
surviving in the superconducting state. Third, in contrast to $U$-based compounds, the
bands at the Fermi energy in $ZrZn_2$ are predominantly $Zr4d$, and magnetism and 
superconductivity result from the same $4d$ electrons.

The observation of superconductivity in $UGe_2,URhGe$ and $ZrZn_2$ suggests that 
superconductivity could occur more generally in ferromagnets. The coexistence of
superconductivity and ferromagnetism may thus be more common and consequently more
important then hitherto realized. 


\section{Theory of ferromagnetic superconductivity}
\subsection{Lattice theory of strongly correlated systems}  

It is generally agreed that the origin of ferromagnetism lies in the Coulomb 
interaction between electrons. The Hamiltonian for electrons with spin $\sigma$
interacting via Coulomb interaction $V(\textbf{r}-\textbf{r}')$ in the presence of ionic
potential $V^{ion}(\textbf{r})$ has the form\cite{ns40}
\begin{equation}
\hat{H}\,=\,\hat{H}_0\,+\,\hat{H}_{int}
\label{ns5}
\end{equation}
where
\begin{equation}
\hat{H}_0\,=\,\sum_{\sigma}\int d^3r \hat{\Psi}_{\sigma}^+(\textbf{r})\left[-\frac {\hbar^2}{2m}\triangle\,+
\,V^{ion}(\textbf{r})\right]\hat{\Psi}_{\sigma}(\textbf{r})
\label{ns6}
\end{equation}
\begin{equation}
\hat{H}_{int}\,=\,\frac 12 \sum_{\sigma\sigma'}\int d^3r\int d^3r' V(\textbf{r}-\textbf{r}') 
\hat{n}_{\sigma}(\textbf{r}) \hat{n}_{\sigma'}(\textbf{r}').
\label{ns7}
\end{equation}
In equations (\ref{ns6}) and (\ref{ns7}) $\hat{\Psi}_{\sigma}(\textbf{r})$, $\hat{\Psi}_{\sigma}^+(\textbf{r})$ 
are electron field operators and 
$\hat{n}_{\sigma}(\textbf{r})=\hat{\Psi}_{\sigma}^+(\textbf{r})\hat{\Psi}_{\sigma}(\textbf{r})$ is the local density
operator. The lattice potential $V^{ion}(\textbf{r})$ leads to a splitting of the parabolic 
dispersion into bands. The non-interacted problem is then characterized by the Bloch wave 
functions $\Phi_{\alpha,\textbf{p}}(\textbf{r})$ and the band energies $\epsilon_{\alpha}(\textbf{p})$, 
where $\alpha$ 
is the band's index. One may introduce Wannier functions localized at the atomic position $\textbf{R}_{i}$ by
the relation
\begin{equation}
\chi_{\alpha,i}\,=\,\frac {1}{\sqrt{N}}\sum_{\textbf{p}}e^{-\imath \textbf{p}\cdot \textbf{R}_i}
\Phi_{\alpha,\textbf{p}}(\textbf{r}),
\label{ns8}
\end{equation} 
where $N$ is the number of lattice sites, and to define the creation and 
annihilation operators $\hat{c}_{\alpha,i\sigma}^+$, $\hat{c}_{\alpha,i\sigma}$ for electrons with
spin $\sigma$ in the band $\alpha$ at site $\textbf{R}_i$ by the equations
\begin{eqnarray}
\hat{c}_{\alpha i\sigma} & = & \int d^3r\chi^{\ast}_{\alpha,i}(\textbf{r})\hat{\Psi}_{\sigma}(\textbf{r})\longleftrightarrow
\hat{\Psi}_{\sigma}(\textbf{r})=\sum_{i\alpha}\chi_{\alpha i}(\textbf{r})\hat{c}_{\alpha i \sigma}.
\label{ns9} \\
\hat{c}^+_{\alpha i\sigma} & = & \int d^3r\chi_{\alpha,i}(\textbf{r})\hat{\Psi}^+_{\sigma}(\textbf{r})\longleftrightarrow
\hat{\Psi}^+_{\sigma}(\textbf{r})=\sum_{i\alpha}\chi^{\ast}_{\alpha i}(\textbf{r})\hat{c}^+_{\alpha i \sigma}.
\label{ns10} 
\end{eqnarray}  
After some algebra the Hamiltonian (\ref{ns5}) may be rewritten in lattice
representation, in terms of creation and annihilation operators 
\begin{equation}
\hat{H}\,=\,\sum_{\alpha ij\sigma}t_{\alpha ij}\hat{c}^+_{\alpha i\sigma}\hat{c}_{\alpha i\sigma}\, 
+\,\frac 12 \sum_{\alpha\beta\gamma\delta}\sum_{ijmn}\sum_{\sigma\sigma'} 
V^{\alpha\beta\gamma\delta}_{ijmn}\hat{c}^+_{\alpha i\sigma}\hat{c}^+_{\beta j\sigma'}\hat{c}_{\delta n\sigma'}
\hat{c}_{\gamma m\sigma},
\label{ns11}
\end{equation}
where the matrix elements are given by
\begin{eqnarray}
t_{\alpha ij} & = & \int d^3r \chi^\ast_{\alpha i}(\textbf{r})\left[-\frac {\hbar^2}{2m}\triangle\,+\,V^{ion}(\textbf{r})\right]
\chi_{\alpha j}(\textbf{r}) \\
\label{ns12a}
V^{\alpha\beta\gamma\delta}_{ijmn} & = & \int d^3r \int d^3r'\chi^{\ast}_{\alpha i}(\textbf{r})
\chi^{\ast}_{\beta j}(\textbf{r}')V(\textbf{r}-\textbf{r}')\chi_{\delta n}(\textbf{r}')\chi_{\gamma m}(\textbf{r}).
\label{ns12}
\end{eqnarray} 

The Hamiltonian (\ref{ns11}) is too general to be tractable. One may restrict the
discussion to a single-band model ($\alpha=\beta=\gamma=\delta=1$).  
The advantage of the single-band model is its comparative mathematical
simplicity. It is simple enough to handle in detail, but yet close enough to
physical realty to supply with useful information, and the obtained effective
model to be of general application. It is physically motivated to discuss this model 
if the Fermi surface lies within a single conduction band, and
if this band is well separated from the other bands and the interaction is
not too strong \cite{ns40}. Accounting for  the weak overlap between neighboring 
orbitals in tight-binding description one may restrict the sums over all sites of
a lattice, to the sum over the nearest neighbors, denoted by $<i,j>$. The remaining 
one-band, nearest-neighbor Hamiltonian has the form\cite{ns40,ns41,ns42} 
\begin{eqnarray}
\hat{H} & = & -t\sum_{<i,j>\sigma}\left(\hat{c}^+_{i\sigma}\hat{c}_{j\sigma}+h.c\right)\,+\,
U\sum_i \hat{n}_{i\uparrow}\hat{n}_{i\downarrow}\,+\,V\sum_{<i,j>}\hat{n}_i \hat{n}_j \nonumber \\
& - & J\sum_{<i,j>}\left(\hat{\textbf{S}}_i\cdot \hat{\textbf{S}}_j\,+\,
\frac 14 \hat{n}_i\hat{n}_j\right)  \\
& + & \sum_{<i,j>}\left[F\left(\hat{c}^+_{i\uparrow}\hat{c}^+_{i\downarrow}
\hat{c}_{j\downarrow}\hat{c}_{j\uparrow}+h.c.\right)\,+\,X\sum_{\sigma}
\left(\hat{c}^+_{i\sigma}\hat{c}_{j\sigma}+h.c\right)
\left(\hat{n}_{i-\sigma}+\hat{n}_{j-\sigma}\right)\right].\nonumber
\label{ns13}
\end{eqnarray}
where $\hat{n}_i=\hat{n}_{i\uparrow}+\hat{n}_{i\downarrow}$ and  
$S^{\mu}_i=1/2\sum_{\sigma\sigma'}\hat{c}^+_{i\sigma}\tau^{\mu}_{\sigma \sigma'}\hat{c}_{i\sigma'}$.
The local(Hubbard) term describes the Coulomb repulsion $(U>0)$, the J-term corresponds 
to the direct Heisenberg exchange which is generically ferromagnetic $(J>0)$ in nature, 
the V-term describes the density-density interaction, the X-term is a density 
dependent hopping and, finally, the F-term describes the hopping of local pairs consisting
of an up and down electron. Coulomb  repulsion $U$ is the largest energy scale in the problem.
In a final step one may neglect the nearest-neighbor interactions 
to obtain the simplest model of strongly correlated electrons, the Hubbard model\cite{ns40}.

\subsection{Magnon-paramagnon effective theory} 

Theories of weak ferromagnetic metals have been developed by several 
theorists \cite{ns43,ns44,ns45,ns46,ns47,ns39,ns48,ns49}. The 
spectrum of the spin excitations has been found. It consists of spin 
fluctuations of paramagnon type and a transverse spin-wave branch. 
The present subsection is devoted to the derivation of an effective 
magnon-paramagnon theory starting from a microscopic single-band 
lattice model of ferromagnetic metals with Hamiltonian
\begin{equation}
\hat H = -t \sum\limits_{<i,j>,\sigma} 
(\hat{c}^+_{i\sigma} \hat{c}_{j\sigma} + {\rm h.c.}) 
-J\sum\limits_{<i,j>} 
\hat{\textbf{S}}_i\cdot \hat{\textbf{S}}_j  
+ U \sum\limits_{i} \hat{n}_{i\uparrow} \hat{n}_{i\downarrow} - 
\mu\sum\limits_i \hat n_i ,
\label{ns14} 
\end{equation} 
where $\mu$ is the chemical potential.

Given that the Hubbard coupling $U$ is the largest energy scale in the theory it
is desirable to diagonalize the Hubbard term. We should also define spin-wave 
excitations such that they are the true Goldstone modes. To accomplish both
of these goals one introduces Schwinger-bosons ($\hat \varphi_{i,\sigma},\hat 
\varphi_{i,\sigma}^{\dagger}$) and slave-fermions ($\hat h_i,\hat 
h_i^{\dagger},\hat d_i,\hat d_i^{\dagger}$) represetation for the operators  
\begin{eqnarray} 
& & \hat {c}_{i\uparrow} = \hat h_i^{\dagger}\hat \varphi _{i1}+ 
\hat \varphi_{i2}^{\dagger}\hat d_i, 
\qquad 
\hat {c}_{i\downarrow} = \hat h_i^{\dagger}\hat \varphi _{i2}- 
\hat \varphi_{i1}^{\dagger}\hat d_i, \nonumber \\ 
& & \hat {c}_{i\uparrow}^{\dagger}\hat {c}_{i\uparrow} 
\hat {c}_{i\downarrow}^{\dagger}\hat {c}_{i\downarrow} = 
\hat d_i^{\dagger}\hat d_i, 
\qquad 
\hat{\vec{S}}_{i}=\frac 12 \sum\limits_{\sigma\sigma'} \hat\varphi^+_{i\sigma} 
{\vec {\tau}}_{\sigma\sigma'} \hat\varphi_{i\sigma'} 
\label{ns15}\\
& & \hat n_i = 1 - \hat h^+_i\hat h_i + \hat d^+_i\hat d_i,\, 
\qquad
\hat \varphi_{i\sigma}^{\dagger}\hat \varphi_{i\sigma}\, 
+\,\hat d_i^{\dagger}\hat d_i\,+\,\hat h_i^{\dagger}\hat h_i\,=\,1 
\nonumber 
\end{eqnarray} 
The partition function can be written as a path integral over the complex 
functions of the Matsubara time $\tau$\,\, $\varphi_{i\sigma}(\tau)\, 
\left(\bar\varphi_{i\sigma}(\tau)\right)$ and Grassmann functions 
$h_i(\tau)\,\left(\bar h_i(\tau)\right)$ and $d_i(\tau)\, 
\left(\bar d_i(\tau)\right)$. 
\begin{equation} 
{\cal Z}(\beta)\,=\,\int\,D\mu\left(\bar\varphi,\varphi,\bar h,h,\bar d,d,\right) 
e^{-S}. 
\label{ns16} 
\end{equation} 
The action is given by the expression 
\begin{equation} 
S=\int\limits^{\beta}_0 d\tau\left[\sum\limits_i\left(\bar\varphi_{i\sigma} 
(\tau) 
\dot\varphi_{i\sigma}(\tau)+\bar h_i(\tau)\dot h_i(\tau)+ 
\bar d_i(\tau)\dot d_i(\tau)\right)+ 
H\left(\bar\varphi,\varphi,\bar h,h,\bar d,d\right)\,\right], 
\label{ns17} 
\end{equation} 
where $\beta$ is the inverse temperature and the Hamiltonian is obtained 
from Eqs.(\ref{ns14}) and (\ref{ns15}) 
replacing the operators with the functions. In terms of the Schwinger 
bosons and slave-fermions the theory is $U(1)$ gauge invariant, and the measure 
includes $\delta$ functions that enforce the constraint  and the 
gauge-fixing condition 

\newpage

\begin{eqnarray} 
& & D\mu\left(\bar\varphi,\varphi,\bar h,h,\bar d,d\right)= 
\prod\limits_{i,\tau,\sigma}\frac {D\bar\varphi_{i\sigma}(\tau) 
D\varphi_{i\sigma}(\tau)}{2\pi i}\prod\limits_{i\tau}D\bar h_i(\tau) 
D h_i(\tau)D\bar d_i(\tau)D d_i(\tau) \nonumber \\ 
& & \prod\limits_{i\tau}\delta\left(\bar\varphi_{i\sigma}(\tau) 
\varphi_{i\sigma}(\tau)\,+\,\bar h_i(\tau) h_i(\tau)\,+\, 
\bar d_i(\tau) d_i(\tau)\,-\,1\right) 
\prod\limits_{i\tau}\delta\left(g.f\right). 
\label{ns18} 
\end{eqnarray}  
  
The ferromagnetic order parameter is a vector field $\textbf{M}$. 
The  transverse spin fluctuations (magnons) are described by ($M_1+iM_2$) and 
($M_1-iM_2$) fields, and the longitudinal  fluctuations (paramagnons) by  
$M_3-<M_3>$. Alternatively the vector field can be written as a product of 
its amplitude $\rho=\sqrt{M_1^2+M_2^2+M_3^2}$ and an unit vector $\textbf{n}$, 
$\textbf{M}=\rho\textbf{n}$. In the ferromagnetic phase one sets $M_3=<M_3>+\varphi$ 
and in linear (spin-wave) approximation obtains $\rho=<M_3>+\varphi$. 
It is evident now, that the fluctuations of the $\rho$ field,  
$\rho-<M_3>$ are exactly the paramagnon excitations in a formalism 
which keeps $0(3)$ symmetry manifest. One can write the 
effective theory in terms of $\textbf{M}$-vector components, or, equivalently, 
in terms of $\rho$ 
and an unit vector $\textbf{n}$. I use the parametrization in 
terms of unite vector and spin singlet amplitude because the unite vector
$\textbf{n}$ describes the true Goldstone modes of the order parameter. 
With that end in view I make a change of variables, introducing new 
Bose fields $f_{i\sigma}(\tau) 
\left(\bar f_{i\sigma}(\tau)\right)$ Ref.\cite{ns50} 
\begin{eqnarray} 
f_{i\sigma}(\tau) & = & \varphi_{i\sigma}(\tau) 
\left[1-\bar h_i(\tau)h_i(\tau)-\bar d_i(\tau)d_i(\tau)\right]^ 
{-\frac 12},\nonumber \\ 
\bar f_{i\sigma}(\tau) & = &\bar \varphi_{i\sigma}(\tau) 
\left[1-\bar h_i(\tau)h_i(\tau)-\bar d_i(\tau)d_i(\tau)\right]^ 
{-\frac 12}, 
\label{ns19} 
\end{eqnarray} 
where the new fields satisfy the constraint 
\begin{equation} 
\bar f_{i\sigma}(\tau)f_{i\sigma}(\tau)\,=\,1. 
\label{ns20} 
\end{equation} 
In terms of the new fields the spin vector has the form 
\begin{equation} 
S^{\mu}_{i}(\tau)=\frac 12 \sum\limits_{\sigma\sigma'}\bar f_{i\sigma}(\tau) 
\tau^{\mu}_{\sigma\sigma'} f_{i\sigma'}(\tau) 
\left[1-\bar h_i(\tau)h_i(\tau)-\bar d_i(\tau)d_i(\tau)\right] 
\label{ns21} 
\end{equation}  
where $n^{\mu}_i\,=\,\sum\limits_{\sigma\sigma'}\bar 
f_{i\sigma}\tau^{\mu}_{\sigma,\sigma'}f_{i\sigma'}$  is a unit vector and 
$\left[1-\bar h_i(\tau)h_i(\tau)-\bar d_i(\tau)d_i(\tau)\right]$ is 
the spin-vector's amplitude.  
When the lattice site is empty or doubly occupied the spin 
vector is zero. When the lattice site is occupied by one electron 
the unit vector  $\textbf{n}_i$  identifies the local orientation. 
One can consider 
the first two components $n_{i1}$ and $n_{i2}$ as 
independent, and then $n_{i3}\,=\,\sqrt{1\,-\,n_{i1}^2\,-\,n_{i2}^2}$. 
In the leading order of the fields, the spin vector has the form      
\begin{equation} 
S_{i1}\simeq \frac 12 n_{i1},  
\qquad 
S_{i2}\simeq \frac 12 n_{i2}, 
\qquad 
S_{i3}\,-\,\frac 12 \simeq -\frac 12 \left(\bar h_i h_i\,+\,\bar d_i 
d_i\right).  
\label{ns22} 
\end{equation}  
The last equation shows that the longitudinal spin fluctuations are associated with 
the collective fields $(\bar h_i h_i\,+\,\bar d_i d_i)$.  
To avoid 
misunderstandings, it is important to point out that the charge-waves are 
associated with the collective field $(\bar d_i d_i\,-\,\bar h_i h_i)$ (see 
the representation of the electron number operator (\ref{ns15})). 

In terms of the new fields the action has the form 
\begin{eqnarray} 
S & = & \int\limits^{\beta}_0 d\tau \left\{\sum\limits_i 
\left[\bar f_{i\sigma}(\tau)\dot f_{i\sigma}(\tau)\,+\, 
\bar h_i(\tau)\left(\frac {\partial}{\partial\tau}-\bar f_{i\sigma}(\tau) 
\dot f_{i\sigma}(\tau) 
\right)h_i(\tau) 
\right.\right. \nonumber \\ 
& & \left.\left.+\,\bar d_i(\tau)\left(\frac {\partial}{\partial\tau}- 
\bar f_{i\sigma}(\tau)\dot f_{i\sigma}(\tau) 
\right)d_i(\tau)\right]\,+\,H\left(\bar f,f,\bar h,h,\bar d,d\right)\right\}, 
\label{ns23} 
\end{eqnarray} 
where $H\left(\bar f,f,\bar h,h,\bar d,d\right)$ is  the Hamiltonian 

To formulate a mean-field theory I drop the terms of order equal or higher 
then six in the Hamiltonian and decouple the four-fermion term, by means 
of the Hubbard-Stratanovich transformation, introducing a real, 
spin-singlet $S_i(\tau)$ field, corresponding to the collective field  
$(\bar h_i h_i\,+\,\bar d_i d_i)$. Now, the action is quadratic 
with respect to the fermions and one can integrate them out. 
The resulting action depends on the spinons $\bar{f}_{i\sigma},f_{i\sigma}$ 
and the real field $S_i$. It has a minimum at the point 
$S_i=s_0, f_{i\sigma}=f_{\sigma}$, and the stationary  condition is  
\begin{equation}  
s_0\,=\,<\bar h_i h_i\,+\,\bar d_i d_i>  
\label{ns24}  
\end{equation}  
Expanding the effective action around the mean field point one obtains 
the effective model in terms of the spinons  
$\bar f_{\sigma}(\tau,\textbf{r}), f_{\sigma}(\tau,\textbf{r})$ and
paramagnons $\varphi_i(\tau)$ 
($2\varphi_i(\tau)=s_0-S_i(\tau)$)\cite{ns51}. 
The first three terms in the expansion have the form
\begin{equation} 
S_{ \textit{eff}}\,=\,S_{\textit{H}}\,+\,S_{\textit{p}}\,+\,S_{\textit{int}}, 
\label{ns25} 
\end{equation} 
 
$S_{\textit{H}}$ is the action of the Heisenberg theory of localized spins. 
In the continuum limit it has the form 
\begin{equation} 
S_{\textit{H}}\,=\,\int\limits^{\beta}_0 d\tau\int d^3r
\left[2M\bar f_{\sigma}(\tau,\textbf{r})\dot 
f_{\sigma}(\tau,\textbf{r})\,+\,\frac {M^2J_r}{2}\sum\limits_{\nu=1}^3 
\partial_{\nu}\textbf{n}(\tau,\textbf{r})\cdot\partial_{\nu}\textbf{n}(\tau,\textbf{r}) 
\right]. 
\label{ns26} 
\end{equation} 
In Eq.(\ref{ns26}),\, $M=\frac 12(1-s_0)$, and $s_0$ comes from 
"tadpoles" diagrams with one $h$ or $d$ line, and the renormalized exchange coupling 
constant $J_r$ is calculated in \cite{ns51}. 

$S_{\textit{p}}$ is the contribution to the effective 
action of the paramagnon excitations   
\begin{equation} 
S_{\textit{p}}\,=\,\frac 12 \int\frac {d\omega}{2\pi}\frac 
{d^3p}{(2\pi)^3}\varphi(\omega,\textbf{p}) 
 \left( r\,+\,a\frac 
{|\omega|}{p}\,+\,b p^2\right)\varphi(-\omega,-\textbf{p}) 
 \label{ns27} 
\end{equation} 
where the constants are obtained from the Lindhard functions for $h$ and $d$ fermions in 
the limit when $p$ and $\frac {\omega}{p}$ are small. 

Finally, the spinon-paramagnon interaction has the form 
\begin{equation} 
S_{\textit{int}}\,=\,M^2\lambda\int\limits^{\beta}_0 d\tau\int d^3r\, \varphi(\tau,\textbf{r}) 
\left[\sum\limits_{\nu =1}^{3}\partial_{\nu}\textbf{n}(\tau,\textbf{r})\cdot 
\partial_{\nu}\textbf{n}(\tau,\textbf{r})\right] 
\label{ns28} 
\end{equation} 
where the effective magnon-paramagnon coupling is obtained from triangular 
diagrams with two $h$ and one $d$ lines or with two $d$ and one $h$ lines.  

To analyze the effective model, it is more convenient to rewrite it in terms of 
rescaled spinon fields  
\begin{equation}  
\bar \zeta_{\sigma}\,=\,\sqrt{2M}\bar f_{\sigma},\qquad 
\zeta_{\sigma}\,=\,\sqrt{2M}f_{\sigma}. 
\label{ns29} 
\end{equation} 
The new fields satisfy the constraint 
\begin{equation} 
\bar \zeta_{\sigma}\zeta_{\sigma}\,=\,2M , 
\label{ns30} 
\end{equation} 
and the action of the effective theory has the form 
\begin{eqnarray}  
S_{ \textit{eff}} & = & \int\limits^{\beta}_0 d\tau\int d^3r
\left[\bar \zeta_{\sigma}(\tau, \textbf{r})
\dot \zeta_{\sigma}(\tau,\textbf{r})\,+\,\frac {J_r}{2}
\sum\limits_{\nu=1}^3  \partial_{\nu}\textbf{M}(\tau,\textbf{r})
\cdot\partial_{\nu}\textbf{M}(\tau,\textbf{r})\right. \nonumber \\ 
& & \left.+\,\frac {\lambda}{4} 
\varphi(\tau,\textbf{r})  \left[\sum\limits_{\nu 
=1}^{3}\partial_{\nu}\textbf{M}(\tau,\textbf{r})\cdot  \partial_{\nu}\textbf{M}(\tau,\textbf{r})\right]\right]\,+\,S_p, 
\label{ns31} 
\end{eqnarray} 
where $\textbf{M}$ is the spin vector 
\begin{equation} 
M^{\mu}\,=\,\frac 12 \bar \zeta_{\sigma}         
\tau^{\mu}_{\sigma,\sigma'}\zeta_{\sigma'}, \qquad 
\textbf{M}^2\,=\,M^2 
\label{ns32} 
\end{equation} 
and $S_p$ is given by Eq.(\ref{ns27}). 
 
It follows from Eq.(\ref{ns21}) that the dimensionless magnetization 
of the system, per lattice site is defined by the equation, 
\begin{equation}  
<S_i^3>=\frac 12 <n_i^3>\left(1-<\bar h_i h_i+ \bar d_i d_i>\right). 
\label{ns33} 
\end{equation}  
At zero temperature $<n_i^3>=1$ and 
using the Eq.(\ref{ns24}) one obtains that $M$ is zero temperature 
dimensionless magnetization of the system per lattice site, $M=<S_i^3>$. The 
parameter $M$ depends on the microscopic parameters of the theory and 
characterizes the vacuum. If, in the vacuum state, every lattice site is 
occupied by one electron with spin up, then $M=\frac 12\quad (s_0=0)$, the 
parameters $a$ and $b$ are equal to zero and $r=\frac {3J}{2}$. In this 
case one can integrate over the paramagnons and the 
resulting theory is the spin $\frac 12$ Heisenberg theory of the localized 
spins. When, in the vacuum state, some of the sites are doubly occupied 
($<\bar d_i d_i>\neq 0$) or empty $(<\bar h_i h_i>\neq 0)$, then 
$M<\frac 12$, the relevant excitations are the spinon and paramagnon 
excitations and the effective theory is a  "spin $M$" Heisenberg  
theory coupled to paramagnon fluctuations defined by  
Eqs.(\ref{ns30},\ref{ns31},\ref{ns32}). The system approaches the 
quantum critical point when $M\rightarrow 0\quad (s_0\rightarrow 1)$, and
$r(M)$ approaches zero when $M\rightarrow 0$. Hence, the parameter $r$ measures the 
distance from the quantum critical point. In quantum paramagnetic phase 
$(M=0)$, the spinon excitations disappear from the spin spectrum (see 
Eqs.(\ref{ns30},\ref{ns32})) and one obtains Hertz's effective model\cite{ns48}. 
(One can add a four-paramagnon term , calculating one-loop diagrams with four $h$ 
or $d$ fermion lines, but I have dropped it motivated by the Hertz's result.) 

In thermal paramagnetic phase (above Curie  
temperature) the spectrum consists of spin singlet fluctuations of paramagnon  
type and spin-$\frac 12$ spinon fluctuations. Well above the  
critical temperature the spinon has a large gap, and the physics of  
ferromagnetic metals is dominated by the paramagnon fluctuations. But just  
above $T_c$ the spinon's gap approaches zero \cite{ns52}, and the  
contribution of the spin-$\frac 12$ fluctuations is essential. 

Below the Curie temperature 
it is convenient to introduce explicitly the magnon excitations 
$a(\tau,\textbf{r}),\,\,\bar a(\tau,\textbf{r})$. To this end, 
I consider the $U(1)$ gauge invariant theory (\ref{ns31}) and impose 
the gauge-fixing condition in the form $arg \zeta_{1}=0$. Then the 
constraint (\ref{ns30}) can be solved by means of the complex field 
$a=\zeta_{2}$ and $\zeta_{1}=\sqrt {2M-\bar a a}$. 
For the components of the spin vector 
$M^+\,=\,M_1\,+\,iM_2,\,\,M^-\,=\,M_1\,-\,iM_2,$ and $M_3$ one obtains the 
Holstein- 
Primakoff representation:   
\begin{eqnarray}  
M^+ & = & \sqrt {2M - \bar a\,a}\,\,a, \,\, 
M^-\,=\,\bar a\sqrt {2M - \bar a\,a}, 
\nonumber \\ 
M^3 & = & M\,-\,\bar a\,a 
\label{ns34} 
\end{eqnarray} 
The kinetic term in the action and the measure are the same as 
the kinetic term and the measure in the theory of a Bose field. 
The only difference is that the complex fields are 
subject to the condition $\bar a\,a\leq 1$. 
 
In the spin-wave theory one approximates $\sqrt {2M-\bar a\,a}$ 
and integrates over the whole complex plane. Then, the model is simplified and 
the effective action can be written in terms of magnon $\bar a,\, a$
and paramagnon $\varphi$ fields 
\begin{eqnarray} 
S_{\textit eff} & = & \int \frac {d\omega}{2\pi}\frac {d^3p}{(2\pi)^3}\left[ 
\bar a(\omega,\textbf{p})\left (i\omega\,+\,\rho p^2\right) a(\omega,\textbf{p})\right.
\nonumber \\
& + & \left.\frac 12 
\varphi(\omega,\textbf{p})\left(r\,+\,a\frac {|\omega|}{p}\,+\,b 
p^2\right)\varphi(-\omega,-\textbf{p})\right]  
\label{ns35} \\
& + & \frac {m\lambda}{2} \int\prod\limits_{l=1}^{2}\frac {d{\omega}_l}{2\pi} 
\frac {d^3p_{l}} 
{(2\pi)^3}\left(\textbf{p}_1\cdot\textbf{p}_2\right)\bar a({\omega}_1,\textbf{p}_1) 
a({\omega}_2,\textbf{p}_2)\varphi({\omega}_1-{\omega}_2,\textbf{p}_1-\textbf{p}_2)
\nonumber 
\end{eqnarray} 
where 
\begin{equation} 
\rho\,=\,M\,J_r 
\label{ns36} 
\end{equation} 
is the spin stiffness constant.  

In the spin-wave approximation the transverse components of the spin 
fields are proportional to the magnon fields  
\begin{equation} 
S^+(\tau,\textbf{r})=\sqrt {2M} a(\tau,\textbf{r}),\,\,\,
S^-(\tau,\textbf{r})=\sqrt {2M} \bar a(\tau,\textbf{r}) 
\label{ns37} 
\end{equation} 
and the field $\varphi(\tau,\textbf{r})$ is exactly the paramagnon (longitudinal 
spin fluctuation)  
\begin{equation} 
S^3(\tau,\textbf{r})-<S^3>\,=\,\varphi(\tau,\textbf{r}). 
\label{ns38} 
\end{equation} 
Hence, in Gaussian approximation, spin-spin 
correlation functions have the form 
\begin{equation} 
D^{\textit{tr}}(\omega,\vec p)\,=\,\frac {2M}{i\omega\,+\,\rho p^2}\,\,\,, 
\qquad\qquad 
D^{\textit{long}}(\omega,\vec p)\,=\,\frac {1}{r\,+ 
a\frac {|\omega|}{p}\, 
+\,bp^2},
\label{ns39} 
\end{equation} 
where the longitudinal magnetic susceptibility is  
\begin{equation} 
\chi\,=\,D^{\textit{long}}(0,0)\,=\,\frac {1}{r} 
\label{ns40} 
\end{equation} 

\subsection{Spin-induced four-fermion interaction}

The next step is to consider model which describes fermions interacting
with their own collective spin fluctuations. The action of the 
effective spin-fermion theory has the form
\begin{eqnarray}
S_{\textit{s-f}} & = & \int\limits^{\beta}_0 d\tau\int d^3r
\left[c^+_{\sigma}(\tau,\textbf{r})
\left(-\frac {1}{2m}\Delta-\mu\right)c_{\sigma}(\tau,\textbf{r})\right. \\
& + & \left.\frac {J}{2} c^+_{\sigma}(\tau,\textbf{r})\tau^{\mu}_{\sigma\sigma'} 
c_{\sigma'}(\tau,\textbf{r})S^{\mu}(\tau,\textbf{r})\right]\,+\,S_{\textit{eff}},
\nonumber
\label{ns41}
\end{eqnarray}
where the second term describes the spin-fermion interaction, and $S_{\textit{eff}}$ 
is the action of the effective magnon-paramagnon theory (\ref{ns35}).
One may represent the spin vector\, $\textbf{S}$\, by means of magnons and paramagnons. 
The partition function  can be then written as a path integral over the complex  
functions of the Matsubara time $\tau$ \,\, 
$a(\tau,\textbf{r}),a^+(\tau,\textbf{r}),\varphi(\tau,\textbf{r})$(magnons, paramagnons) and Grassmann
functions  $c_{\sigma}(\tau,\textbf{r}),c^+_{\sigma}(\tau,\textbf{r})$  
\begin{equation}
{\cal Z}(\beta)\,=\,\int D\mu\left(a^+,a,\varphi,c^+_{\sigma},c_{\sigma}\right)
e^{-S_{\textit{s-f}}}.
\label{ns42}
\end{equation}
In the spin-wave approximation Eqs.(\ref{ns37},\ref{ns38}) the effective action\, 
$S_{\textit{s-f}}$\, is quadratic with respect to the spin fluctuations, and one may 
integrate them out using the formula for the Gaussian integral 
\cite{ns53}. As a result one obtains an effective Fermi theory. It is convenient
to write the action as a sum\, $S_{\textit{f}}=S_{\textit{f}-0}+S_{\textit{f-int}}$
of a free part 
\begin{equation}
S_{\textit{f}-0}\,=\,\int\limits^{\beta}_0 d\tau\int \frac {d^3p}{(2\pi)^3}
\left[c^+_{\sigma}(\tau,\textbf{p})\dot c_{\sigma}(\tau,\textbf{p})
+\epsilon_{\sigma}(p) c^+_{\sigma}(\tau,\textbf{p})c_{\sigma}(\tau,\textbf{p}),
\right],
\label{ns43}
\end{equation}
where $\epsilon_{\sigma}(p)$ are the dispersions of spin-up and spin-down fermions
in ferromagnetic phase. 
\begin{equation}   
\epsilon_{\uparrow}(p)= \frac {p^2}{2m}-\mu-\frac {JM}{2},  
\quad \,\,\,
\epsilon_{\downarrow}(p)=\frac {p^2}{2m}-\mu +\frac {JM}{2},  
\label{ns44}  
\end{equation}    
and four-fermion interaction resulting from the exchange of spin fluctuations
\begin{eqnarray} 
& & S_{\textit{f-int}}\,=\,-\frac {J^2}{8}\int
d^4x_1\int d^4x_2\left[c_{\uparrow}^+(x_1)c_{\uparrow}(x_1)
- c_{\downarrow}^+(x_1)c_{\downarrow}(x_1)\right] \nonumber \\
& & D^{\textit{long}}(x_1-x_2) 
\left[c_{\uparrow}^+(x_2)c_{\uparrow}(x_2)-c_{\downarrow}^+(x_2)c_{\downarrow}(x_2)\right] \\
\label{ns45} 
& & -\,\frac {J^2}{4}\int d^4x_1d^4x_2
c_{\downarrow}^+(x_1)c_{\uparrow}(x_1)D^{\textit{tr}}(x_1-x_2) 
c_{\uparrow}^+(x_2)c_{\downarrow}(x_2). \nonumber
\end{eqnarray} 
where $x=(\tau,\textbf{r})$, $D^{\textit{long}}$ is the paramagnon propagator, 
and $D^{\textit{tr}}$ is the magnon Green function Eq.(\ref{ns39}). 

For the purpose of doing analytical calculations it is convenient to approximate 
the four-fermion interaction with the static one. To this end I replace the magnon 
and paramagnon propagators Eq.(\ref{ns39} by static potentials 
\begin{eqnarray}
D^{\textit{tr}}(\omega,\textbf{p}) & \rightarrow & V_{m}(p)=\frac {2M}{\rho p^2} \\
\label{ns46}
D^{\textit{long}}(\omega,\textbf{p}) & \rightarrow & V_{pm}(p)=\frac {1}{r+b\,p^2}.
\nonumber
\end{eqnarray} 

Let us represent the spin anti-parallel composite field 
$c_{\uparrow}c_{\downarrow}$ as a sum of symmetric and antisymmetric parts. 
After some algebra one obtains an effective four 
fermion theory which can be written as a sum of four terms. Three of them 
describe the interaction of the components of spin-1 composite fields 
$(\uparrow\uparrow,
\uparrow\downarrow+\downarrow\uparrow,\downarrow\downarrow)$. 
The fourth term describes the
interaction of the spin singlet composite fields  
$\uparrow\downarrow-\downarrow\uparrow$. The Hamiltonians of interactions are  
\begin{equation}  
H_{\uparrow\uparrow}\,=\,-\frac {J^2}{8}\int\prod\limits_{i}\frac  
{d^3p_i}{(2\pi)^3}  
\left[c_{\uparrow}^+(\textbf{p}_1)c_{\uparrow}^+(\textbf{p}_2)  
 c_{\uparrow}(\textbf{p}_2-\textbf{p}_3)
c_{\uparrow}(\textbf{p}_1+\textbf{p}_3)\right]  
 V_{pm}(\textbf{p}_3)
\label{ns47}  
\end{equation} 
\begin{equation}  
H_{\downarrow\downarrow}\,=\,-\frac {J^2}{8}\int\prod\limits_{i}\frac  
{d^3p_i}{(2\pi)^3}  
\left[c_{\downarrow}^+(\textbf{p}_1)c_{\downarrow}^+(\textbf{p}_2) 
c_{\downarrow}(\textbf{p}_2-\textbf{p}_3)c_{\downarrow}(\textbf{p}_1+\textbf{p}_3)\right]  
 V_{pm}(\textbf{p}_3) 
\label{ns48}  
\end{equation}  
\begin{eqnarray}  
& & H_{\uparrow\downarrow+\downarrow\uparrow} = -\frac 
{J^2}{8}\int\prod\limits_{i}\frac  {d^3p_i}{(2\pi)^3}  
\left[c_{\uparrow}^+(\textbf{p}_1)c_{\downarrow}^+(\textbf{p}_2)+  
c_{\downarrow}^+(\textbf{p}_1)c_{\uparrow}^+(\textbf{p}_2)\right] \nonumber\\ 
& & \times\left[c_{\uparrow}(\textbf{p}_2-\textbf{p}_3)c_{\downarrow}(\textbf{p}_1+\textbf{p}_3)+  
c_{\downarrow}(\textbf{p}_2-\textbf{p}_3)c_{\uparrow}(\textbf{p}_1+\textbf{p}_3)\right]  
V_{-}(\textbf{p}_3)  
\label{ns49} 
\end{eqnarray}  
\begin{eqnarray}  
& & H_{\uparrow\downarrow-\downarrow\uparrow,} = \frac 
{J^2}{16}\int\prod\limits_{i}\frac  {d^3p_i}{(2\pi)^3}  
\left[c_{\uparrow}^+(\textbf{p}_1)c_{\downarrow}^+(\textbf{p}_2)-  
c_{\downarrow}^+(\textbf{p}_1)c_{\uparrow}^+(\textbf{p}_2)\right] \nonumber \\  
& & \times\left[c_{\uparrow}(\textbf{p}_2-\textbf{p}_3)c_{\downarrow}(\textbf{p}_1+\textbf{p}_3)-  
c_{\downarrow}(\textbf{p}_2-\textbf{p}_3)c_{\uparrow}(\textbf{p}_1+\textbf{p}_3)\right]  
V_{+}(\textbf{p}_3), 
\label{ns50} 
\end{eqnarray}  
where  
\begin{equation}  
V_{-}(p)=\frac {2M}{\rho p^2}\,-\,\frac {1}{r+b p^2},\qquad   
V_{+}(p)=\frac {2M}{\rho p^2}\,+\,\frac {1}{r+b p^2} .  
\label{ns51}  
\end{equation} 

The spin singlet fields' interaction Eq.(\ref{ns50}) 
is repulsive and does not contribute to the superconductivity\cite{ns54}.  
The spin parallel fields' interactions Eqs.
(\ref{ns47},\ref{ns48}) are mediated by the exchange of longitudinal
spin fluctuations and the resulting state is spin-parallel Cooper pairing.  
The interaction of the $\uparrow\downarrow+\downarrow\uparrow$ fields 
Eq.(\ref{ns49}) is relevant for magnon-induced superconductivity. 
It has an attracting part due to exchange of magnons and a 
repulsive part due to  exchange of paramagnons.   
 
\subsection{Paramagnon exchange mechanism of \\ FM-superconductivity}

The most popular theory of FM-superconductivity is based on the paramagnon
exchange mechanism\cite{ns55,ns56} with Hamiltonians of interaction 
Eqs.(\ref{ns47}) and (\ref{ns48}).
By means of the Hubbard-Stratanovich transformation one introduces  
$\uparrow\uparrow$ and $\downarrow\downarrow$
composite fields, the order parameters, and then the fermions 
can be integrated out. The obtained free energy is a function 
of the composite fields and the integral over the composite fields can be 
performed approximately by means of the steepest descend method. To this end 
one sets the first derivatives of the free energy with respect to composite 
fields equal to zero, these are the gap equations. To obtain the critical
temperature $T_{sc}$ one considers the finite temperature gap equations
linearized at critical temperature\cite{ns56}
\begin{equation}
\Delta_{\sigma}(\textbf{p})\,=\,\frac {1}{2Z^2(0)}\int \frac {d^3k}{(2\pi)^3}
V_{pm}(\textbf{p}-\textbf{k})\frac {\tanh(\frac {\beta_{sc}}{2}
\epsilon^{\ast}_{\sigma}(\textbf{k}))}{\epsilon^{\ast}_{\sigma}(\textbf{k})}
\Delta_{\sigma}(\textbf{k}).
\label{ns52}
\end{equation}
where $Z(0)$ is the mass renormalization constant $m^\ast/m=Z(0)$.
For nearly ferromagnetic system, it scales with magnetization $M$ like 
$Z(0)\sim\ln(1/M)$\cite{ns57}. In Eq.(\ref{ns52}), 
$\epsilon^{\ast}_{\sigma}(\textbf{p})$ are the spin-${\sigma}$ dispersions (\ref{ns44})
with $m\rightarrow m^{\ast}$, and $\beta_{sc}$ is the inverse critical 
temperature. To solve the gap equations, one expands the gap functions in
spherical harmonics $Y_{lm}$ and keeps only the $l=1,m=0$ component. Then,
the superconducting critical temperature can be obtained following McMillan
approximation
\begin{equation}
T_{sc}^{\sigma}\,=\,1.14\omega(M)\exp[-Z(0)/\lambda^{\sigma}_1],
\label{ns53}
\end{equation} 
where
\begin{equation}
\lambda^{\sigma}_1\,=\,\frac {N^{\sigma}(0)}{2(k_f^{\sigma})^2}
\int\limits_0^{2k_f^{\sigma}} 
dk k \left (1-\frac {k^2}{2(k_f^{\sigma})^2}\right)V_{pm}(k),
\label{ns54}
\end{equation}
$k_f^{\sigma}$ are the Fermi wavevectors for the spin-up and spin-down Fermi
serfaces, and $N^{\sigma}(0)$ is the density of states at the spin-$\sigma$
Fermi surface. Near the quantum phase transition ($M=0$), $\lambda^{\sigma}_1$
diverges like $\ln(1/M)$ and cancels the logarithmic singularity of the
renormalization parameter $Z(0)$. As a result, the M dependence of the critical
temperature is determined by $\omega(M)$. The superconducting 
transition temperatures as a function of the exchange interaction 
parameter $\overline{I}$ are depicted in Fig.5. In the ferromagnetic phase 
$\overline{I}>1$, and the magnetic transitions occurs at $\overline{I}=1$.  
The parameter $r=\overline{I}-1 $ measures the distance from the quantum 
critical point (see Eqs.(\ref{ns39},\ref{ns40})) and scales with magnetization
like $r\sim M^2$\cite{ns39,ns44,ns47,ns51}.    
\begin{center}
\begin{figure}[htb]
\label{nsfig5}
\centerline{\psfig{file=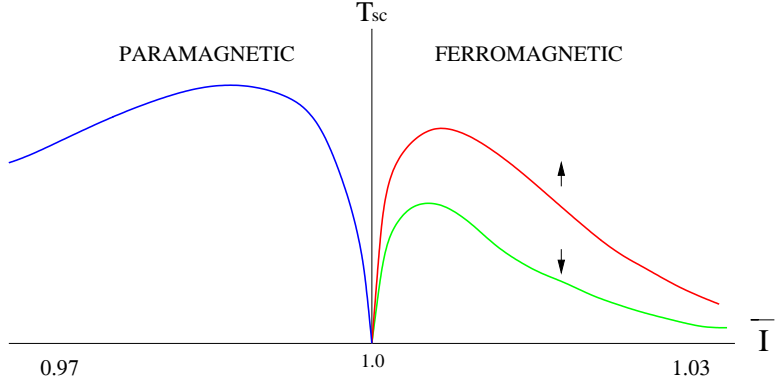,width=11cm,height=5cm}}
\caption{Paramagnon exchange mechanism of superconductivity.The superconducting 
transition temperature as a function of the exchange interaction 
parameter $\overline{I}$.}
\end{figure}
\end{center}

\vspace*{-0.7cm}

In paramagnetic phase the spectrum consists of spin singlet fluctuations of paramagnon  
type. The exchange of paramagnons leads to a four-fermion interaction, which is  
attractive in triplet channel. 
The rotational symmetry, in continual limit, requires all three components 
$(\uparrow\uparrow,\uparrow\downarrow+\downarrow\uparrow,\downarrow\downarrow)$  
of the spin-1 vector to be nonequal to zero. The superconducting 
transition temperature as a function of the exchange interaction 
parameter $\overline{I}$ is depicted in Fig.5. In the paramagnetic phase 
$\overline{I}<1$, and
the parameter $r=1-\overline{I}$ measures the distance from the quantum 
critical point.

The superconducting critical temperature in Fay and Appel
theory increases when the magnetization decreases and very close to the
quantum critical point falls down rapidly Fig.5.  It has recently been the subject
of controversial debate. It is obtained in\cite{ns58}, by means of a more
complete Eliashberg treatment, that the transition temperature is nonzero at
the critical point. In \cite{ns59},  however, the authors have shown that the
reduction of quasiparticle coherence and life-time due to spin fluctuations is
the pair-breaking  process which leads to a rapid reduction of the
superconducting critical temperature near the quantum critical point. 
In order to explain the absence of superconductivity in paramagnetic phase 
of $UGe_2$, $URhGe$ and $ZrZn_2$ it was accounted for the magnon
paramagnon interaction and proved that the critical temperature is much
higher in the ferromagnetic phase than in the paramagnetic one\cite{ns60}.

Despite of the efforts, the improved theory of paramagnon induced
superconductivity can not cover the whole variety of properties of 
FM superconductivity. 

\subsection{Magnon induced FM superconductivity}

Magnon exchange mechanism of superconductivity has been proposed \cite{ns61} 
to explain in a natural way the fact that the  
superconductivity in $UGe_2$, $ZrZn_2$ and  
$URhGe$ is confined to the ferromagnetic  
phase.The order parameter is a spin anti-parallel component  
$\uparrow\downarrow+\downarrow\uparrow$ of  
a spin-1 triplet  $(\uparrow\uparrow, 
\uparrow\downarrow+\downarrow\uparrow, \downarrow\downarrow)$ 
with zero spin projection.
The effective Hamiltonian of the system is   
\begin{equation}  
H_{\textit{eff}}=H_0+H_{\uparrow\downarrow+\downarrow\uparrow}  
\label{ns55}  
\end{equation}  
where  $H_0$ is the Hamiltonian of the free spin  
up and spin down fermions with dispersions Eq.(\ref{ns44}) and Hamiltonian
of interaction (\ref{ns49}). 
The transverse spin fluctuations are pair forming and the 
longitudinal ones are pair breaking.

By means of the Hubbard-Stratanovich transformation one introduces 
composite field  
$\uparrow\downarrow+\downarrow\uparrow$ and then the fermions 
can be integrated out. The integral over the composite field can be 
performed approximately by means of the steepest descend method. To this end 
one sets the derivative of the free energy with respect to composite 
field equal to zero, this is the gap equation. To ensure that the 
fermions which form Cooper pairs are the same as those responsible for 
spontaneous magnetization, one has to consider the equation for the 
magnetization as well.
\begin{equation}  
M=\frac 12 <c^+_{\uparrow}c_{\uparrow}-c^+_{\downarrow}c_{\downarrow}> 
\label{ns56} 
\end{equation} 
The system of equations for the gap and for the magnetization
determines the phase where the superconductivity and the ferromagnetism
coexist.

The system can be written in terms of Bogoliubov excitations, which have 
the following dispersions relations:
\begin{eqnarray} 
& & E_1(\textbf{p}) = -\frac 
{JM}{2}-\sqrt{\epsilon^2(\textbf{p})+|\Delta(\textbf{p})|^2} \nonumber \\  
& & E_2(\textbf{p}) = \frac 
{JM}{2}-\sqrt{\epsilon^2(\textbf{p})+|\Delta(\textbf{p})|^2}   
\label{ns57} 
\end{eqnarray} 
where $\Delta(\textbf{p})$ 
is the gap, and $\epsilon(\textbf{p})=\frac {p^2}{2m}-\mu$.  
At zero temperature the equations take the form 
\begin{eqnarray} 
M & = & \frac 12\int\frac {d^3k}{(2\pi)^3}\left[1-\Theta (-E_2(\textbf{k}))\right] 
\label{ns58}\\ 
\Delta(\textbf{p}) & = & \frac {J^2}{8}\int\frac {d^3k}{(2\pi)^3}\,\frac {V(\textbf{p}-\textbf{k})
\,\,\Theta (-E_2(\textbf{k}))}{\sqrt {\epsilon^2(\textbf{k})+
|\Delta(\textbf{k})|^2}}\,\Delta (\textbf{k}) 
\label{ns58a} 
\end{eqnarray} 

The gap is an antisymmetric function $\Delta (-\textbf{p})=-\Delta (\textbf{p})$, so 
that the expansion in terms of spherical harmonics $Y_{lm}(\Omega_{\textbf{p}})$ 
contains only terms with odd $l$. I assume that the component with $l=1$ and 
$m=0$ is nonzero and the other ones are zero
\begin{equation} 
\Delta (\textbf{p})=\Delta_{10}(p)\sqrt {\frac {3}{4\pi}}\cos\theta. 
\label{ns59} 
\end{equation}
Expanding the potential $V_{-}(\textbf{p}-\textbf{k})$ in terms of Legendre polynomial $P_l$ 
one obtains that only the 
component with $l=1$ contributes the gap equation. The potential $V_1(p,k)$ 
has the form, 
\begin{eqnarray} 
V_{1}(p,k) & = & \frac {3M}{\rho}\left[\frac 
{p^2+k^2}{4p^2k^2}\ln\left(\frac {p+k}{p-k}\right)^2-\frac{1}{pk}\right] 
 \nonumber \\ 
& - &  
\frac {3M}{\rho}\beta \left[\frac {p^2+k^2}{4p^2k^2}\ln\frac 
{r'+(p+k)^2}{r'+(p-k)^2}\,-\,\frac {1}{pk}\right],  
\label{ns60} 
\end{eqnarray} 
where $3M/\rho=3/\rho_0$, $\beta=\rho/2Mb=\rho_0/2b>1$\, 
and $r'=r/b<<1$. A straightforward analysis  
shows that for a fixed $p$ , the potential is positive when $k$ 
runs an interval around $p$ $(p-\Lambda,p+\Lambda)$, 
where $\Lambda$ is approximately independent on $p$.  
In order to allow for an explicit analytic solution, I introduce further 
simplifying assumptions by neglecting the dependence of $\Delta_{10}(p)$ 
on $p$ ($\Delta_{10}(p)=\Delta_{10}(p_f)=\Delta$) and setting 
$V_1(p_f,k)$ equal to a constant $V_1$ within interval 
$(p_f-\Lambda,p_f+\Lambda)$ and zero elsewhere. The system of equations 
(\ref{ns58},\ref{ns58a})  
is then reduced to the system
\begin{eqnarray} 
M & = & \frac 
{1}{8\pi^2}\int\limits_0^{\infty}dkk^2\int\limits_{-1}^{1}dt[1-\Theta(-E_2(k,t))] 
\label{ns61}\\ 
\Delta & = & \frac {J^2V_1}{32\pi^2}\int\limits_{p_f-\Lambda}^{p_f+\Lambda} dk 
k^2\int\limits_{-1}^{1}dt\,t^2\frac 
{\Theta(-E_2(k,t))}{\sqrt{\epsilon^2(k)+\frac {3}{4\pi}t^2\Delta^2}} 
\Delta 
\label{ns61a} 
\end{eqnarray} 
where $t=\cos\theta$.

\subsubsection{Solution which satisfies $\sqrt {\frac {3}{\pi}}\Delta<JM$}

The equation of magnetization 
(\ref{ns61}) shows that it is convenient to represent the gap in the form  
$\Delta= \sqrt {\frac {\pi}{3}}\kappa (M) JM$, where $\kappa (M)<1$. 
Then the equation
\begin{equation} 
E_2(p,t)=0, 
\label{ns62}
\end{equation}
defines the Fermi surfaces,
\begin{equation} 
p^{\pm}_{f}=\sqrt {p^2_{f}\pm m\sqrt{J^2M^2-\frac 
{3}{\pi}t^2\Delta^2}}\,\,,\,\,\, p_{f}=\sqrt {2\mu m} 
\label{ns63} 
\end{equation}  

The domain between the Fermi surfaces contributes to the magnetization $M$ in 
Eq.(\ref{ns61}), but it is cut out from the domain of integration in the gap 
equation Eq.(\ref{ns61a}). When the magnetization increases, the 
domain of integration in the gap equation decreases. Near the quantum 
critical point the size of the gap is small, and hence the linearized gap 
equation can be considered. Then it is easy to obtain the critical value of 
the magnetization $M_{SC}$ \cite{ns62}.

When the magnetization approaches 
zero, the domain between the Fermi surfaces decreases. One can approximate the 
equation for magnetization Eq.(\ref{ns61}) substituting $p^{\pm}_{f}$ from 
Eq.(\ref{ns63}) in the 
the difference $(p^{+}_{f})^2-(p^{-}_{f})^2$ and setting 
$p^{\pm}_{f}=p_{f}$ elsewhere. Then, in this approximation, the 
magnetization is linear in $\Delta$, namely 
\begin{equation} 
\Delta =\sqrt {\frac {\pi}{3}}J\kappa M 
\label{ns64} 
\end{equation} 
where $\kappa$ runs the interval $(0,1)$, and satisfies the equation
\begin{equation} 
\kappa\sqrt {1-\kappa^2}+\arcsin\kappa\,=\frac {8\pi^2}{mp_{f}J} 
\label{ns65} 
\end{equation} 
The Eq.(\ref{ns65}) has a solution if $mp_{f}J>16\pi$. Substituting $M$ from 
Eq.(\ref{ns64}) in Eq.(\ref{ns61a}), one arrives at an equation for the gap.
This  equation can be solved in a standard way and the solution is
\begin{eqnarray} 
\Delta & = & \sqrt {\frac {16\pi}{3}}\frac {\Lambda p_{f}\kappa}{m} 
\exp\left[-\frac32 
I(\kappa)-\frac {24\pi^2}{J^2 V_1 m p_{f}}\right] 
\label{ns66} \\ 
I(\kappa) & = & \int\limits_{-1}^{1}dt t^2 \ln\left(1+\sqrt 
{1-\kappa^2 t^2}\right) \nonumber 
\end{eqnarray} 
Eqs (\ref{ns64},\ref{ns65},\ref{ns66}) are the solution of the system 
Eqs.(\ref{ns61},\ref{ns61a}) near the quantum transition to paramagnetism.  
\begin{center}
\begin{figure}[htb]
\label{nsfig6}
\centerline{\psfig{file=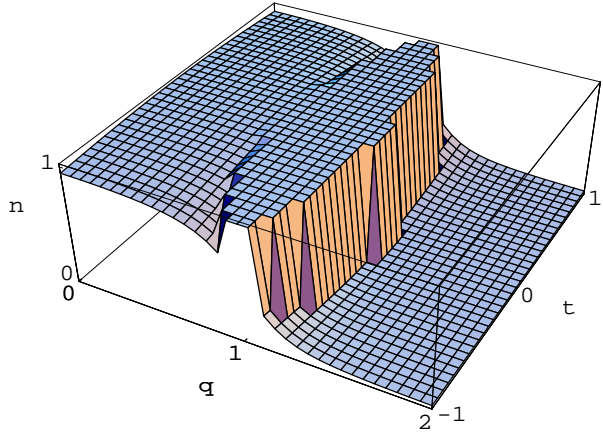,width=8cm,height=6cm}}
\caption{The zero temperature momentum distribution $n$, for spin up  
fermions, as a function of $q=\frac {p}{p_{f}}$ and $t=\cos\theta$.}
\end{figure}
\end{center}

\vspace*{-0.7cm}

One can write the momentum 
distribution functions $n^{\uparrow}(p,t)$ and $n^{\downarrow}(p,t)$ of the 
spin-up and spin-down quasiparticles 
in terms of the distribution functions of the Bogoliubov fermions  
\begin{eqnarray} 
n^{\uparrow}(p,t) & = & u^2(p,t)n_1(p,t)+v^2(p,t)n_2(p,t) 
\label{ns67} \\ 
n^{\downarrow}(p,t) & = & u^2(p,t)(1-n_1(p,t))+v^2(p,t)(1-n_2(p,t)) 
\nonumber 
\end{eqnarray} 
where $u(p,t)$ and $v(p,t)$ are the coefficients in the Bogoliubov 
transformation. At zero temperature $n_1(p,t)=1$, 
$n_2(p,t)=\Theta(-E_2(p,t))$, and the Fermi surfaces Eq.(\ref{ns62}) manifest 
themselves both in the spin-up and spin-down momentum distribution functions. 
The functions are depicted in Fig.6 and Fig.7.
\begin{center}
\begin{figure}[htb]
\label{nsfig7}
\centerline{\psfig{file=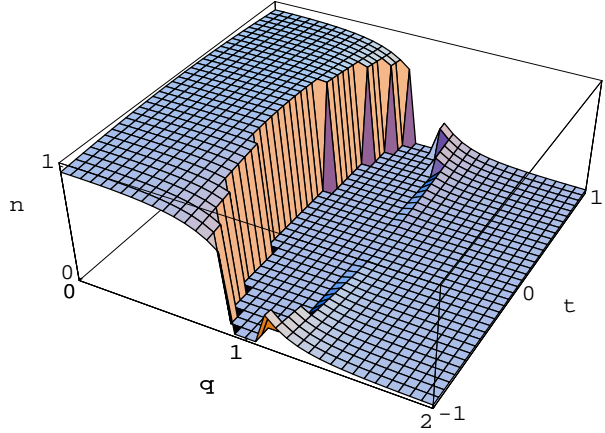,width=8cm,height=6cm}}
\caption{The zero temperature momentum distribution $n$, for spin-down  
fermions, as a function of $q=\frac {p}{p_{f}}$ and $t=\cos\theta$.}
\end{figure}
\end{center}

\vspace*{-0.7cm}

The two Fermi surfaces explain the mechanism of Cooper pairing.   
In the ferromagnetic phase $n^{\uparrow}$ and $n^{\downarrow}$ have different  
(majority and minority) Fermi surfaces (see Fig.8 and Fig.9, $t=0$ graphs).  
\begin{center}
\begin{figure}[htb]
\label{nsfig8}
\centerline{\psfig{file=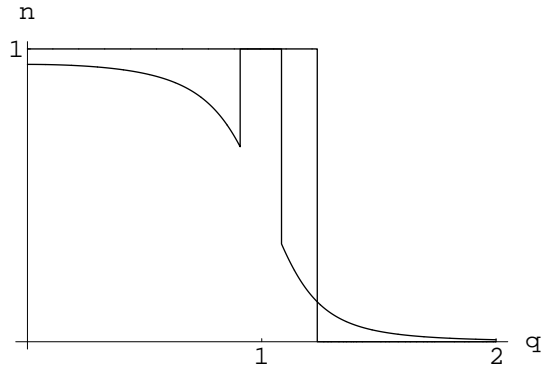,width=8cm,height=5cm}}
\caption{The zero temperature momentum distribution $n$, for spin-up  
fermions, as a function of $q=\frac {p}{p_{f}}$ for $t=0$ (the gap is zero)
and $t=\pm$ (the gap is maximal).}
\end{figure}
\end{center}

\vspace*{-0.7cm}

\begin{center}
\begin{figure}[htb]
\label{nsfig9}
\centerline{\psfig{file=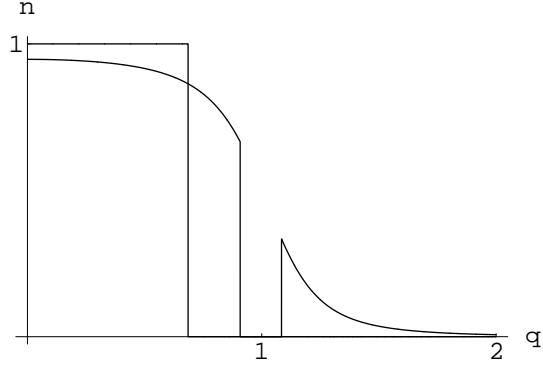,width=8cm,height=5cm}}
\caption{The zero temperature momentum distribution $n$, for spin-down 
fermions, as a function of $q=\frac {p}{p_{f}}$ for $t=0$ (the gap is zero)
and $t=\pm$ (the gap is maximal).}
\end{figure}
\end{center}

\vspace*{-0.7cm}

The spin-up electrons contribute the majority 
Fermi surface, and spin-down electrons contribute the minority Fermi surface. 
When the value of the momentum of the emitted or absorbed magnon lies within  
interval $(p_f-\Lambda,p_f+\Lambda)$ the effective potential between spin-up and  
spin-down electrons is attracting. Hence, if the Fermi momenta $p_f^{\uparrow}$ and 
$p_f^{\downarrow}$ lie within interval $(p_f-\Lambda,p_f+\Lambda)$ the interaction 
between spin-up electrons, which contribute the majority Fermi surface, and spin-down  
electrons, which contribute the minority Fermi surface, is attracting. As a result,  
spin-up electrons from majority Fermi surface transfer to the minority Fermi surface 
and form spin anti-parallel Cooper pairs, while spin-down electrons from minority 
Fermi surface transfer to the majority one and form spin anti-parallel Cooper pairs too. 
As a result, the onset of superconductivity is accompanied by the appearance of a  
second Fermi surface in each of the spin-up and spin-down momentum  
distribution functions (see Fig.8 and Fig.9, $t=1$ graphs). 

The existence of the two Fermi surfaces explains the linear dependence 
of the specific heat at low temperatures:
\begin{equation} 
\frac {C}{T}\,=\,\frac {2\pi^2}{3}\left(N^+(0)+N^-(0)\right) 
\label{ns68} 
\end{equation} 
Here $N^{\pm}(0)$ are the density of states on the Fermi surfaces. 
One can rewrite the $\gamma=\frac {C}{T}$ constant in terms of Elliptic 
Integral of the second kind $E(\alpha,x)$ 
\begin{equation} 
\gamma\,=\,\frac {m p_{f}}{3\kappa}
\left[(1+s)^{\frac 12} E(\frac 12 \arcsin\kappa,\frac 
{2s}{s+1})+  
(1-s)^{\frac 12} E(\frac 12 \arcsin\kappa,\frac 
{2s}{s-1})\right],
\label{ns69} 
\end{equation} 
where $s=JMm/p^2_{f}<1$ and  
$\kappa=\sqrt {3/\pi}\,\Delta/JM$. 

\subsubsection{Solution which satisfies $\sqrt {\frac {3}{\pi}}\Delta>JM$}

In the present sub-chapter one
looks for a solution of the system which satisfies
\begin{equation}
\sqrt {\frac {3}{\pi}}\Delta\,>\,JM
\label{ns70}
\end{equation} 
The inequality Eq.(\ref{ns70})
shows that the gap can not be arbitrarily small when the magnetization
is finite. Hence the system undergoes the quantum phase transition from
ferromagnetism to FM-superconductivity with a jump. Approaching the quantum
critical point from the ferromagnetic side, one sets the gap equal to zero
in the equation for the magnetization (\ref{ns61}) and considers the gap 
equation (\ref{ns61a}) with magnetization as a parameter. It is more 
convenient to consider the free energy as a function of the gap for the
different values of the parameter $M$. To this purpose I introduce the
dimensionless "gap" $x$ and the parameters $s,\lambda$ and $g$ 
\begin{equation} 
x=\sqrt {\frac {3}{\pi}}\frac {m}{p_{f}^2}\Delta,\,\, 
s=\frac {m}{p_{f}^2}JM,\,\,\lambda=\frac {\Lambda}{p_{f}},\,\, 
g=\frac {J^2V_1mp_{f}}{8\pi^2}  
\label{ns71} 
\end{equation} 
Then the free energy is a function of $x$ 
and depends on the parameters $s, \lambda$ and $g$.
The dimensionless free energy $F(x)$ is depicted in Fig.10 for
$\lambda=0.08,\,g=20$ and three values of the parameter $s$, $s=0.8, s=0.69$
and $s_{cr}=0.595$.
\begin{eqnarray} 
& & F(x)=\frac {6m^2}{\pi p_{f}^4}\left({\cal F}(x)-{\cal F}(0)\right) = x^2+  
g\int\limits_{1-\lambda}^{1+\lambda}dq q^2\int\limits_{-1}^1dt\,\times\nonumber 
\\ & & \left[\left(s-\sqrt {(q^2-1)^2+t^2x^2}\right) 
\Theta(\sqrt{(q^2-1)^2+t^2x^2}-s)-\right. \nonumber \\ 
& & \left. \left(s-\sqrt {(q^2-1)^2}\right) 
\Theta(\sqrt{(q^2-1)^2}-s)\right] 
\label{ns72} 
\end{eqnarray} 
\begin{center}
\begin{figure}[htb]
\label{nsfig10}
\centerline{\psfig{file=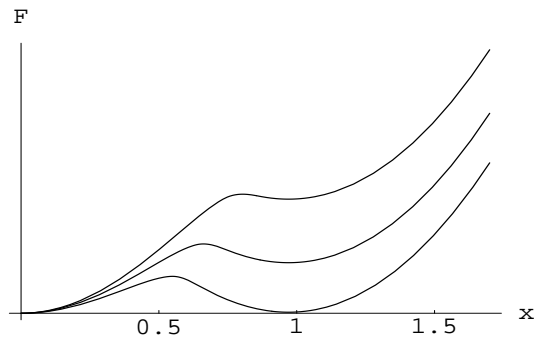,width=8cm,height=4.5cm}}
\caption{The dimensionless free energy $F(x)$ as a function of 
dimensionless gap $x$. $\lambda=0.08,\,g=20$, $s_1=0.8$(upper line), 
$s_2=0.69$(middle line) and $s_{cr}=0.595$(lower line).}
\end{figure}
\end{center}

\vspace*{-0.7cm}

As the graph shows, for some values of the microscopic
parameters $\lambda$ and $g$, and decreasing the parameter $s$ (the
magnetization), the system passes trough a first order quantum phase
transition. The critical values $s_{cr}$ and $x_{cr}$ satisfy 
$x_{cr}/s_{cr}=\sqrt {3/\pi}\,\Delta_{cr}/JM_{cr}>1$ in
agreement with Eq.(\ref{ns70}).

Varying the microscopic parameters beyond the critical values, one has to solve 
the system of equations (\ref{ns61},\ref{ns61a}). One represents again the gap in the form
\begin{equation}  
\Delta= \sqrt {\frac {\pi}{3}}\kappa (M) JM
\label{ns73}
\end{equation} 
but now $\kappa (M)>1$.
Then the equation $E_2(k,t)=0$, which defines the Fermi surface, has no 
solution if 
$-1<t<-1/\kappa (M)$ and $1/\kappa(M)<t<1$, 
and has two solutions
\begin{equation} 
p^{\pm}_{f}=\sqrt {p^2_{f}\pm m\sqrt{J^2M^2-\frac 
{3}{\pi}t^2\Delta^2}} 
\label{ns74} 
\end{equation}  
when $-1/\kappa(M)<t<1/\kappa (M)$. 

The solutions (\ref{ns74}) determine the two pieces of the Fermi surface. They stick 
together at $t=\pm\,1/\kappa(M)$, so that the Fermi surface is simple 
connected. The domain between pieces contributes to the 
magnetization $M$ in Eq.(\ref{ns61}), but it is cut out from the domain of 
integration in the gap equation Eq.(\ref{ns61a}). The Fermi surface manifests itself 
both in the spin-up and spin-down momentum distribution functions. 
The functions are depicted in Fig.11 and Fig.12.
\begin{center}
\begin{figure}[htb]
\label{nsfig11}
\centerline{\psfig{file=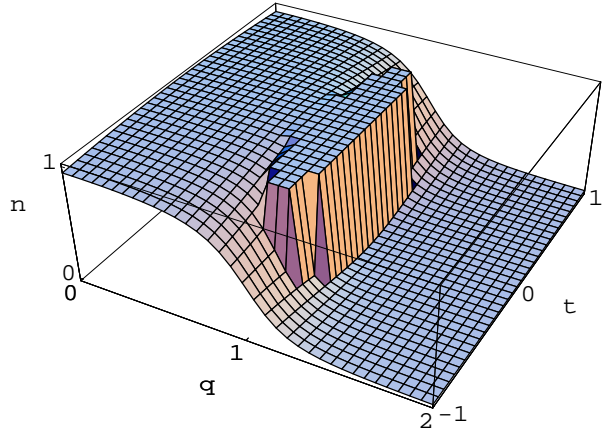,width=8cm,height=6cm}}
\caption{The zero temperature momentum distribution $n$, for spin up  
fermions, as a function of $q=\frac {p}{p_{f}}$ and $t=\cos\theta$.}
\end{figure}
\end{center}

\vspace*{-0.7cm}

When the magnetization approaches zero, one can approximate the  
equation for magnetization Eq.(\ref{ns61}) substituting $p^{\pm}_{f}$ from  
Eq.(\ref{ns74}) in the  
the difference $(p^{+}_{f})^2-(p^{-}_{f})^2$ and setting  
$p^{\pm}_{f}=p_{f}$ elsewhere. Then, in this approximation, the  
magnetization is linear in $\Delta$, namely  
\begin{equation}  
\Delta =\sqrt {\frac {\pi}{3}}J\kappa M  
\label{ns75}  
\end{equation}  
where $\kappa=\frac {mp_{f}J}{16\pi}$ is the small magnetization limit of  
$\kappa(M)$.  
The Eq.(\ref{ns75}) is a solution if $mp_{f}J>16\pi$ (see Eq.(\ref{ns70})).  
Substituting $M$ from  
Eq.(\ref{ns75}) in Eq.(\ref{ns61a}), one arrives at an equation for the gap. 
This equation can be solved in a standard way and the 
solution is 
\begin{equation}  
\Delta\,=\,\sqrt {\frac {16\pi}{3}}  
\frac {p_{f}\Lambda}{m}  
\exp \left[-\frac {24\pi^2}{mp_{f}J^2V_1}-\frac {\pi}{4\kappa^3}+\frac  
{1}{3}\right]  
\label{ns76}  
\end{equation}  
Eqs (\ref{ns75},\ref{ns76}) are the solution of the system  
near the quantum transition to paramagnetism.   
The second derivative of the free energy Eq.(\ref{ns72}) with respect  
to the gap is positive when $mp_{f}J/16\pi>(21\pi/16)^{1/3}$,  
hence the state where the  
superconductivity and the ferromagnetism coexist is stable.
\begin{center}
\begin{figure}[htb]
\label{nsfig12}
\centerline{\psfig{file=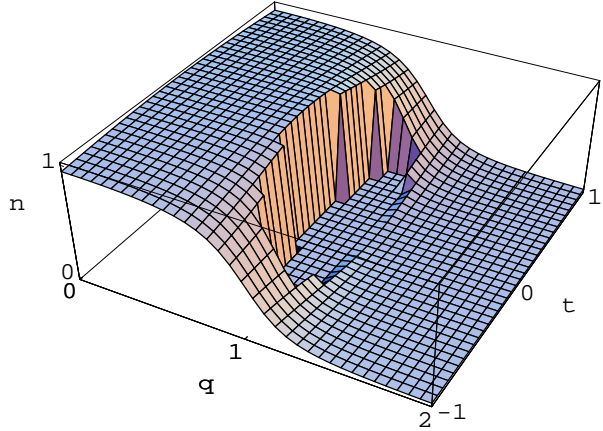,width=8cm,height=6cm}}
\caption{The zero temperature momentum distribution $n$, for spin-down  
fermions, as a function of $q=\frac {p}{p_{f}}$ and $t=\cos\theta$.}
\end{figure}
\end{center}

The existence of the Fermi surface explains the linear dependence  
of the specific heat at low temperature:   
\begin{equation}  
\frac {C}{T}\,=\,\frac {2\pi^2}{3}N(0)  
\label{ns77}  
\end{equation}  
Here $N(0)$ is the density of states on the Fermi surface.  
One can rewrite the $\gamma=C/T$ constant in terms of Elliptic  
Integral of the second kind $E(\alpha,x)$  
\begin{equation}
\gamma\,=\,\frac {m p_{f}}{3\kappa(M)} 
\left[(1+s)^{\frac 12} E(\frac {\pi}{4},\frac  
{2s}{s+1})+ 
(1-s)^{\frac 12} E(\frac {\pi}{4},\frac {2s}{s-1})\right],  
\label{ns78}  
\end{equation}
where $s<1$.
Eq.(\ref{ns78}) shows that for $\kappa(M)$ just above one the specific heat 
constant $\gamma$  is smaller in ferromagnetic phase, while for $\kappa(M)>>1$
it is smaller in FM-superconducting phase.The result closely matches the 
experiments with $ZrZn_2$ and $URhGe$ respectively.  

The solutions Eqs.(\ref{ns64},\ref{ns73})show that magnetization and
superconductivity disappear simultaneously. It results from the equation of
magnetization, which in turn is added to ensure that the 
fermions which form Cooper pairs are the same as those responsible for 
spontaneous magnetization. Hence, the fundamental assumption that
superconductivity and ferromagnetism are caused by the same electrons leads
to the experimentally observable fact that the quantum phase transition is a
transition to paramagnetic phase without superconductivity.

An important experimental fact is that $ZrZn_2$ and $URhGe$ are
superconductors at ambient pressure as opposed to the existence of a quantum
phase transition in $UGe_2$. To comprehend this difference one
considers the potential (\ref{ns60}). The quantum phase transition results from
the existence of a momentum cutoff $\Lambda$, above which the potential is
repulsive. In turn, the cutoff excistence follows from the relation
$\beta=\rho/2Mb>1$, which is true when the spin-wave approximation
expression for the spin stiffness constant $\rho=M\rho_0$ is used. The spin
wave approximation correctly describes systems with a large magnetization, for
example $UGe_2$. But in order to study systems with small magnetization, one
has to account for the magnon-magnon interaction which changes the small
magnetization  asymptotic of $\rho$, $\rho=M^{1+\alpha}\rho_0$, where
$\alpha>0$. Then for a small $M$ $\beta<1$, and the potential is attractive
for all momenta. Hence for systems which, at ambient pressure, are close to
quantum critical point, as $ZrZn_2$ and $URhGe$, the magnon self-interaction
renormalizes the spin fluctuations parameters so that the magnons dominate
the pair formation and quantum phase transition can not be observed. But if
one applies an external magnetic field, the magnon opens a gap proportional
to the magnetic field. Increasing the magnetic field the paramagnon
domination leads to first order quantum phase transition.  
   
The proposed model of ferromagnetic superconductivity differs from the
models discussed in \cite{ns55,ns56} in many aspects. First, the
superconductivity is due to the exchange of magnons, and the model describes
in an unified way the superconductivity in $UGe_2,\,ZrZn_2$ and $URhGe$.
Second, the paramagnons have
pair-breaking effect. So, the understanding the mechanism of paramagnon
suppression is crucial in the search for the ferromagnetic superconductivity
with higher critical temperature. For example, one can build such a bilayer
compound that the spins in the two layers are oriented in two
non-collinear directions, and the net ferromagnetic moment is nonzero.
The paramagnon in this phase is totally suppressed and the low lying excitations
consist of magnons and additional spin wave modes with linear dispersion
$\epsilon(p)\sim p$\cite{ns63}. If the new spin-waves are pair breaking, their
effect is weaker than those of the paramagnons, and hence the superconducting
critical temperature should be higher. 
Third, the order parameter is a
spin antiparallel component of a spin triplet with zero spin projection.
The existence of two Fermi surfaces in each of the spin-up and spin-down
momentum distribution functions 
leads to a linear temperature dependence of the specific heat at low
temperature.

The proposed model of magnon-induced superconductivity does not contain the
relativistic effects, namely spin-orbit coupling which is present in $UGe_2$.
The resulting magneto-crystalline anisotropy will modify the spin-wave
excitation and will add a gap in the magnon spectrum, which changes the
potential Eq.(\ref{ns60}).

\section{Conclusions}

The brief review involves a personal choice of topic and emphasis. Here
I have concentrated on the coexistence of ferromagnetism and 
superconductivity. The main experimental conclusion is that 
the same band electrons 
are subject to the ferromagnetic and superconducting instability. The review
is devoted to spin exchange mechanism of superconductivity. I have not
discussed the phonon mechanism of triple p-type superconductivity\cite{ns64,ns64a},
the symmetry and nodal structure of the order parameter\cite{ns65,ns66} and other. 

The physics of ferromagnetic superconductivity involves a subtle interplay between
magnetism and superconductivity. 
The concept of magnon induced superconductivity is set within the general scheme of
itinerant magnetism. It is suggested by the fact that superconducting phase lies 
entirely in the ferromagnetic one. The theory explains the ferromagnetic to 
FM-superconducting quantum phase transition and the absence of specific heat anomaly.
A point which requires further theoretical investigation is the presence of an 
additional phase line within ferromagnetic phase.  
There is no an adequate theoretical explanation of the resistivity anomaly near 
the characteristic temperature $T_x$. Further theoretical work is needed 
to understand why $T_x$ decreases with pressure and disappears at a pressure
$p_x$ close to the pressure at which the superconductivity is strongest. 

\vspace*{0.6cm}
\begin{center}
{\large {\bf Acknowledgements}}
\end{center}
\vspace*{0.3cm}
 
The author would like to thank C. Pfleiderer, T.M.Rice and C. Honerkamp for valuable 
discussions. This research is supported by the Sofia University  
Science Foundation Grant-2003.


\end{document}